\documentclass[aps,pre,showpacs,groupedaddress,superscriptaddress,twocolumn]{revtex4}
\usepackage{amsmath}
\usepackage{amssymb}
\usepackage{latexsym}

\newcommand{\be}{\begin{equation}}
\newcommand{\ee}{\end{equation}}
\newcommand{\br}{\begin{eqnarray}}
\newcommand{\er}{\end{eqnarray}}

\newcommand{\bd}{\begin{displaymath}}
\newcommand{\ed}{\end{displaymath}}

\newcommand{\bfig}{\begin{figure}}
\newcommand{\efig}{\end{figure}}
\input{tcilatex}
\begin{document}

\title{Quantum master equations from classical Lagrangians with two
stochastic forces}
\author{A. V. Dodonov}
\email{adodonov@df.ufscar.br}
\author{S. S. Mizrahi}
\email{salomon@df.ufscar.br} \affiliation{Departamento de
F\'{\i}sica, CCET, Universidade Federal de S\~{a}o Carlos, Via
Washington Luiz km 235, 13565-905, S\~ao Carlos, S\~ao Paulo,
Brazil}
\author{V. V. Dodonov}
\email{vdodonov@fis.unb.br} \affiliation{Instituto de F\'{\i}sica,
Universidade de Bras\'{\i}lia, PO Box 04455, 70910-900,
Bras\'{\i}lia, Distrito Federal, Brazil}
\date{\today }

\begin{abstract}
We show how a large family of master equations, describing quantum Brownian
motion of a harmonic oscillator with translationally invariant damping, can
be derived within a phenomenological approach, based on the assumption that
an environment can be simulated by two classical stochastic forces. This
family is determined by three time-dependent correlation functions (besides
the frequency and damping coefficients), and it includes as special cases
the known master equations, whose dissipative part is bilinear with respect
to the operators of coordinate and momentum.
\end{abstract}

\maketitle

\section{introduction}

\label{introduction}

The phenomena of irreversibility and damping in quantum systems have been
the subjects of numerous studies from the very first years of Quantum
Mechanics (QM) \cite{Land27,Bloch28,Pauli28,WW30}. They are attributed to
the action of some environment that drains information from quantum system $%
\mathcal{S}$, thus increasing the entropy during its evolution. The
environment is usually referred as reservoir $\mathcal{R}$ whose variables
are expressed as $q$-numbers (operators acting in the Hilbert space), and it
is assumed to have an infinity of degrees of freedom. In the most general
case, one has to treat the whole quantum system $\mathcal{S}+\mathcal{R}$,
taking into account all details of each subsystem and the interaction
between them. It is known, however, that under certain conditions the
influence of environment can be described approximately by means of a few
parameters which enter some dynamical equations containing only the
variables related to the system $\mathcal{S}$. 
%
In the Schr\"{o}dinger picture such equations for the statistical operator $%
\hat{\rho}(t)$ or its different representations (density matrix, Wigner
function, etc.), known under the name \emph{quantum master equations\/}
(QME), were studied in numerous papers, e.g., \cite%
{vanHove57,MonShu57,Zwan58,Toda58,George60,Fain62,zwanzig64,LouWal65,Opp65,Uller66,Weid,Lax,LouiMar67,Glaub-42,LambScul,Ful68,peier,bripimi,Gall96,Bre01,Haba01}%
. An alternative description in the Heisenberg picture is achieved within
the frameworks of \emph{quantum Langevin equations\/} or Langevin-Heisenberg
equations (LHE's) \cite%
{Lax,Sen,Schwin61,Haus,FordKac65,Baus67,hermiz,lindenberg,GaCol85,PimRaz87,Ford88,Strat97,Kan051,D05,Kal06}%
. These two approaches were discussed in detail in many books and reviews
\cite%
{Agar73,Haake73,louisell,Dav,Pen79,Spohn,Dek,Talk86,Grab88,Carm93,Gard00,Breu02,Dbook03}%
.

The shortest and simplest \emph{formal\/} way of introducing relaxation in
quantum mechanics is to postulate some general structure of the master or
Langevin--Heisenberg equations, which guarantees fulfilment of the
fundamental requirements of QM (such as the properties of hermiticity and
positivity of the statistical operator, as well as its normalization; or
preservation of canonical commutation relations between the time-dependent
operators) for any instant of time. For example, a general structure of
dynamical mappings of density operators preserving positive semidefiniteness
was established in \cite{JorSud61,Kraus71}, and equivalent differential
equations were considered in \cite{Bausch66,Belav,Gori76,Lind76}. For a more
restricted problem of \emph{quantum Brownian motion\/} this approach was
used, e.g., in Refs. \cite{DOM85,167,Sand87,Adam99,Vacc02,Brod04}, where
some sets of free parameters were chosen in such a way that \emph{mean
values\/} of coordinates and momenta obeyed given classical equations of
motion.

However, it seems desirable to have some schemes permitting to \emph{derive\/%
} master equations from some general principles. The most natural way to do
this is to start from some explicit Hamiltonian for the total system $%
\mathcal{S}+\mathcal{R}$. Then the dynamical equations for the subsystem $%
\mathcal{S}$ arise as a result of taking the trace over the reservoir
variables. In this approach both the subsystem and reservoir are considered
as quantum objects from the very beginning. A frequently used simple model
of the reservoir consists of an infinite number of harmonic oscillators
(HO's) with a frequency distribution chosen according to some hypothesis.
Note, however, that although the use of a reservoir is usually viewed as a
\textquotedblleft microscopic\textquotedblright\ approach for the derivation
of LHE or QME, in reality, unless one knows \emph{precisely\/} the nature of
the environment and its interaction with $\mathcal{S}$ (described by a
Hamiltonian with parameters dependent on fundamental constants; see, e.g.,
\cite{Ford85} for this exceptional case), strictly speaking, it should be
considered as \emph{phenomenological\/}. This occurs because several
assumptions must be made concerning the very nature of the reservoir modes,
on the parameters of the reservoir Hamiltonian and on the $\mathcal{R-S}$
interaction term.

Taking this point of view, it seems interesting to study, what kinds of
master equations can be obtained, if one follows the phenomenological path
from the very beginning to the very end. This is the aim of our paper. As a
general principle we assume the Lagrange--Hamiltonian formalism, which
impelled the development of physics in general and in particular QM and
field theories. It is complemented by the assumption that $\mathcal{S}$
interacts with time-dependent (TD) classical fluctuating forces [$F_{i}(t)$%
], which simulate the action of the environment without any further
preoccupation about the nature of their microscopic origin. We show that
combining some phenomenological Lagrangian, which takes into account the
effect of dissipation, with a suitably chosen set of classical stochastic
forces and following standard rules of the \textquotedblleft canonical
quantization\textquotedblright\ procedure, one can obtain in a quite
straightforward and simple manner a large class of master equations, which
embraces many equations, introduced earlier within the frameworks of
different schemes. This approach was used earlier in Refs.\cite%
{Pap73,svinin,dekker2,brimiz,AlMes82}. However, in those papers the authors
considered only \emph{one\/} classical stochastic force. As a result, the
equations obtained had some restrictions on their domain of validity. We
show that introducing \emph{two\/} stochastic forces one can obtain more
general equations.

The paper is organized as follows. In Sec. \ref{section-GME} we give a brief
review of known master equations describing quantum damped oscillators. New
results are contained in Sec. \ref{section-EFME}, where we derive the master
equation with the use of two effective classical fluctuating forces. We show
that in the classical case, it is always possible to use a single force,
without changing the physics. However, in the quantum case two forces are
necessary to obtain equations compatible with the principles of QM. Sec. \ref%
{section-summary} contains a summary and conclusions. In appendix A we
expose some details of calculations.

\section{Conventional master equations}

\label{section-GME}

The Hamiltonian $H_0$ considered in this paper corresponds to a particle of
mass $m$ and angular frequency $\omega _{0}$ subjected to a one-dimensional
harmonic force,
\begin{equation}
\hat{H}_{0}=\hbar \omega _{0}\left( \hat{a}^{\dagger }\hat{a}+1/2\right) ={%
\hat{p}^{2}}/(2m) + {m\omega _{0}^{2}}\hat{x}^{2}/2,  \label{Hosc}
\end{equation}
where the non-hermitian lowering operator $\hat{a}$ can be written in terms
of hermitian operators, the particle position $\hat{x}$ and its momentum $%
\hat{p}$, as
\begin{equation*}
\hat{a}=\left( 2\hbar \omega _{0} m\right) ^{-1/2}\left( {m}\omega _{0}\hat{x%
}+{i\hat{p}}\right).
\end{equation*}

The general form of the master equation is
\begin{equation}
{d\hat{\rho}}/{dt}+(i/\hbar )\left[ H,\hat{\rho}\right] =\mathcal{L}\hat{\rho%
},  \label{ssme1}
\end{equation}%
where the dissipative nonunitary superoperator $\mathcal{L}$ accounts for
the influence of the environment and the operator $H$ can be different from $%
H_0$. One of the most frequently used master equations for the damped
harmonic oscillator has an origin in the problems of quantum optics. It
corresponds to $H=H_0$ and
\begin{eqnarray}
\mathcal{L}\hat{\rho} &=&{\lambda }\left( \bar{n}+1\right) \left( 2\hat{a}%
\hat{\rho}\hat{a}^{\dagger }-\hat{a}^{\dagger }\hat{a}\hat{\rho}-\hat{\rho}%
\hat{a}^{\dagger }\hat{a}\right)  \notag \\
&&+{\lambda }\bar{n}\left( 2\hat{a}^{\dagger }\hat{\rho}\hat{a}-\hat{a}\hat{a%
}^{\dagger }\hat{\rho}-\hat{\rho}\hat{a}\hat{a}^{\dagger }\right) ,
\label{SME}
\end{eqnarray}%
where 
$\bar{n}\equiv \bar{n}(\omega _{0},T)=(e^{\beta }-1)^{-1}$
is the reservoir mean number of quanta at temperature $T$ ($\beta =\hbar
\omega _{0}/k_{B}T$), $k_{B}$ is Boltzmann constant and $\lambda $ is the
decay rate or damping constant, which contains the strength parameter of the
interaction between $\mathcal{R}$ and $\mathcal{S}$. The operator (\ref{SME}%
) seems to be derived for the first time under the assumption of \emph{weak
coupling} between a quantum system $\mathcal{S}$ and a quantum \emph{thermal}
reservoir and by adopting the Born-Markov approximation in \cite%
{Weid,Lax,LouiMar67,Glaub-42}. Its representation in the Fock basis was
derived in the general form in \cite{Toda58,George60,Fain62,LambScul},
although the special case of the equation for \emph{diagonal matrix
elements\/} can be traced to papers by Landau, Bloch and Pauli \cite%
{Land27,Bloch28,Pauli28} (see also \cite{MonShu57}).

An immediate generalization of operator (\ref{SME}) is
\begin{equation}
\mathcal{L}\hat{\rho}=\sum_{j}\left( 2\hat{\Phi}_{j}\hat{\rho}\hat{\Phi}%
_{j}^{\dag }-\hat{\Phi}_{j}^{\dag }\hat{\Phi}_{j}\hat{\rho}-\hat{\rho}\hat{%
\Phi}_{j}^{\dag }\hat{\Phi}_{j}\right) ,  \label{Lind}
\end{equation}%
where $\hat{\Phi}_{j}$ are arbitrary linear operators (their number also may
be arbitrary). Sometimes the right-hand side of (\ref{Lind}) is called the
\textquotedblleft Lindblad form\textquotedblright , after the study \cite%
{Lind76}, although this general structure, preserving the hermiticity,
normalization and positivity of $\hat{\rho}$, was discovered by several
authors earlier 
\cite{Bausch66,Belav,Gori76}.

The relaxation operator in terms of operators $\hat{x}$ and $\hat{p}$ is
usually associated with the problem of \emph{quantum Brownian motion\/}. The
most general master equation, preserving the normalization and hermiticity
of the statistical operator $\hat\rho$ and containing only \emph{bilinear\/}
forms of operators $\hat{x}$ and $\hat{p}$, corresponds to the choice
\begin{equation}
H = H_0 + \frac{\mu}{2} \left\{ \hat{x},\hat{p}\right\},  \label{Hmu}
\end{equation}
\begin{eqnarray}
\mathcal{L}\hat{\rho} &=& \frac{i{\lambda }}{2\hbar }\left[ \hat{p},\left\{
\hat{x},\hat{\rho}\right\} \right] -\frac{i{\lambda }}{2\hbar }\left[ \hat{x}%
,\left\{ \hat{p},\hat{\rho}\right\} \right] -\frac{D_{p}}{\hbar ^{2}}\left[
\hat{x},\left[ \hat{x},\hat{\rho}\right] \right]  \notag \\
&& -\frac{D_{x}}{\hbar ^{2}}\left[ \hat{p},\left[ \hat{p},\hat{\rho}\right] %
\right] +\frac{D_{z}}{\hbar ^{2}}\left[ \hat{x},\left[ \hat{p},\hat{\rho}%
\right] \right] +\frac{D_{z}}{\hbar ^{2}}\left[ \hat{p},\left[ \hat{x},\hat{%
\rho}\right] \right],  \label{ssme2}
\end{eqnarray}%
where $\mu$, $\lambda$, $D_x$, $D_p$ and $D_z$ can be, in principle,
arbitrary functions of time. This form was considered, e.g., in Refs. \cite%
{Baus67,Dek,Sand87} (in the case of time-independent coefficients) and \cite%
{Kan051} (with arbitrary time-dependent coefficients). Some differences in
explicit expressions can be removed with the aid of identities $[\hat{x},%
\hat{p}]=i\hbar$ and
\begin{equation*}
\left[ \hat{x},\left\{ \hat{p},\hat{\rho}\right\} \right] +\left[ \hat{\rho}%
,\left\{ \hat{x},\hat{p}\right\} \right] +\left[ \hat{p},\left\{ \hat{\rho},%
\hat{x}\right\} \right] =0.
\end{equation*}%
The meaning of parameters $\lambda $ and $\mu $ becomes clear from the
equations for mean values of the coordinate and momentum:
\begin{eqnarray}
d\langle \hat{x}\rangle /dt &=&\langle \hat{p}\rangle /m+(\mu -\lambda
)\langle \hat{x}\rangle ,  \label{eqmeanx} \\
d\langle \hat{p}\rangle /dt &=&-m\omega _{0}^{2}\langle \hat{x}\rangle -(\mu
+\lambda )\langle \hat{p}\rangle .  \label{eqmeanp}
\end{eqnarray}%
The choice ${\lambda }={\mu }$ eliminates the friction term from the
equation for $d\langle \hat{x}\rangle /dt$. This special case was studied by
Dekker in \cite{dekker2}. Generalizations of (\ref{Hmu}) and (\ref{ssme2})
to multidimensional systems (in particular, to the case of a charged
particle in a magnetic field) were given in \cite{DOM85,167}.  The
superoperator (\ref{SME}) is a particular case of (\ref{ssme2}) for
\begin{equation*}
D_{p}=\frac{\lambda }{2}m\hbar \omega _{0}\left( \bar{n}+\frac{1}{2}\right)
=(m\omega _{0})^{2}D_{x},\quad D_{z}=\mu =0.
\end{equation*}

The operator master equation (\ref{ssme2}), being written in terms of the
Wigner function,
\begin{equation}
W\left( x,p,t\right) =\frac{1}{\pi \hbar }\int dye^{-2ipy/\hbar
}\left\langle x-y|\hat{\rho}(t)|x+y\right\rangle ,  \label{wigner0}
\end{equation}%
assumes a simple form of the Fokker--Planck equation,
\begin{eqnarray}
\frac{\partial W}{\partial t} &=&\frac{\partial }{\partial p}\left( \left[
m\omega _{0}^{2}x+(\mu +\lambda )p\right] W\right)  \notag \\
&&-\frac{\partial }{\partial x}\left( \left[ p/m+(\mu -\lambda )x\right]
W\right)  \notag \\
&&+D_{p}\frac{\partial ^{2}W}{\partial p^{2}}+D_{x}\frac{\partial ^{2}W}{%
\partial x^{2}}+2D_{z}\frac{\partial ^{2}W}{\partial p\partial x}.
\label{FP}
\end{eqnarray}%
Thus we see that the terms proportional to $D_{x}$, $D_{p}$, and $D_{z}$ in
Eq. (\ref{ssme2}) describe the diffusion in the phase space.

Introducing the phase space vector variable $\mathbf{q}=(x,p)$, one can
write Eq. (\ref{FP}) in a compact form
\begin{equation}
\frac{\partial W\left( \mathbf{q},t\right) }{\partial t}=-\frac{\partial }{%
\partial q_{i}}\left[ \left( \mathbf{A}\mathbf{q}\right) _{i}W\right] +D_{ij}%
\frac{\partial ^{2}W}{\partial q_{i}\partial q_{j}}  \label{wigner}
\end{equation}%
(sum over repeated indices is understood), where
\begin{equation}
\mathbf{A}=\left\Vert
\begin{array}{cc}
\mu -\lambda  & m^{-1} \\
-m\omega _{0}^{2} & -(\mu +\lambda )%
\end{array}%
\right\Vert   \label{A}
\end{equation}%
is a `drift' matrix, which governs the evolution of the first order
statistical moments (mean values),
\begin{equation}
d\langle \mathbf{q}\rangle /dt=\mathbf{A}\langle \mathbf{q}\rangle .
\label{dot-q-A}
\end{equation}%
Introducing the covariances $\sigma _{jk}=\frac{1}{2}\langle \hat{q}_{j}\hat{%
q}_{k}+\hat{q}_{k}\hat{q}_{j}\rangle -\langle \hat{q}_{j}\rangle \langle
\hat{q}_{k}\rangle $, one can verify that both equations, (\ref{FP}) and (%
\ref{wigner}), result in the following equation for the symmetrical
covariance matrix $\mathbf{M}\equiv \Vert \sigma _{jk}\Vert $:
\begin{equation}
{d\mathbf{M}}/{dt}=\mathbf{A}\mathbf{M}+\mathbf{M}\mathbf{\tilde{A}}+2%
\mathbf{D},  \label{wigner3}
\end{equation}%
where $\mathbf{\tilde{A}}$ is the transposed matrix of $\mathbf{A}$ and $%
\mathbf{D}\equiv \Vert D_{ij}\Vert $ is the symmetrical diffusion matrix ($%
D_{12}=D_{21}=D_{z}$).

Taking the operators $\hat\Phi_j$ in Eq. (\ref{Lind}) as linear combinations
of operators $\hat{x}$ and $\hat{p}$,
\begin{equation}
\hat\Phi_j = \alpha_j \hat{x} + \beta_j \hat{p},  \label{Phi-xp}
\end{equation}
one can verify that the operator (\ref{ssme2}) can be rewritten in the form (%
\ref{Lind}), provided the following conditions can be satisfied \cite%
{Zven-osc}:
\begin{equation}
\sum_j |\beta_j|^2 = D_x /\hbar^2, \quad \sum_j |\alpha_j|^2 = D_p /\hbar^2,
\label{DxDp-ab}
\end{equation}
\begin{equation}
\sum_j \alpha_j^* \beta_j = i\lambda/(2\hbar) - D_z /\hbar^2.
\label{Dz-ab}
\end{equation}
In view of the Schwartz inequality,
\begin{equation*}
\sum_j |\beta_j|^2 \sum_j |\alpha_j|^2 \ge |\sum_j \alpha_j^* \beta_j|^2,
\end{equation*}
the condition of compatibility of Eqs. (\ref{DxDp-ab}) and (\ref{Dz-ab}) is
the inequality
\begin{equation}
D_{p}D_{x}-D_{z}^{2}\geq (\hbar \lambda /2)^{2} \equiv [\hbar\mbox{Tr}(%
\mathbf{A})/4]^2.  \label{restrDxpz}
\end{equation}
The condition (\ref{restrDxpz}), which was derived and discussed from
different points of view in Refs. \cite%
{Dbook03,DOM85,167,Sand87,Adam99,Vacc02,Zven-osc,Barch,Vals}, guarantees
that the positivity of statistical operator is preserved for all times and
\emph{for any physically admissible initial state}. This is the necessary
and sufficient condition (together with conditions $D_x \ge 0$ and $D_p \ge 0
$) of reducibility of the operator (\ref{ssme2}) to the Lindblad form (\ref%
{Lind}) \cite{DOM85,167}. Note that parameter $\mu $ does not enter
the constraint (\ref{restrDxpz}), because it is related to the
correction to the Hamiltonian (\ref{Hmu}) and not to the non-unitary
part of the total Liouville
superoperator. So its presence in the ``friction forces'' in Eqs. (\ref%
{eqmeanx}) and (\ref{eqmeanp}) is not relevant to the existence or
non-existence of the Lindblad representation of the master equation.

Some frequently considered master equations with \emph{time-independent
coefficients\/}, such as, e. g., the Agarwal equation \cite{Agar71} (with $%
\lambda =\mu $, $D_{p}=2m\lambda \omega _{0}\bar{n}$ and $D_{x}=D_{z}=0$) or
its special case, known as the Caldeira--Leggett equation \cite{cl} (with $%
\lambda =\mu $, $D_{p}=2m\lambda k_{B}T$ and $D_{x}=D_{z}=0$), do not
satisfy the condition (\ref{restrDxpz}). Consequently, these equations can
result in violations of the positivity of the statistical operator (which is
equivalent to the violation of the uncertainty relations \cite%
{Talk86,167,AlMes82,Zven-osc}) at the intermediate stages of evolution, if
one tries to use them outside the domain of their validity (which
corresponds to the limit of high temperatures, $k_{B}T\gg \hbar \omega _{0}$%
).

On the other hand, it was shown in \cite{haake,UnZu89,Hu92,DJRLR95,Zer95}
that the Wigner function of a subsystem $\mathcal{S}$ interacting with a
reservoir $\mathcal{R}$ satisfies Eq. (\ref{wigner}) \emph{at any time\/},
if, (I) the total Hamiltonian of the whole system $\mathcal{S}+\mathcal{R}$
is an arbitrary quadratic form with respect to coordinates and momenta (in
particular, the interaction Hamiltonian can be an arbitrary bilinear form
with respect to the coordinates of $\mathcal{S}$ and $\mathcal{R}$) and if,
(II) the initial statistical operator of the total system is factorized, $%
\hat{\rho}_{tot}=\hat{\rho}_{\mathcal{S}}\hat{\rho}_{\mathcal{R}}$, where $%
\hat{\rho}_{\mathcal{R}}$ is an arbitrary \emph{Gaussian\/} state (i.e., not
necessarily thermal, it can be \emph{squeezed\/}, for example). However, in
such a case, (a) the matrices $\mathbf{A}$ and $\mathbf{D}$ are \emph{%
explicitly time-dependent\/} (which is interpreted sometimes as a
manifestation of \emph{non-Markovian evolution\/} \cite{Kan051,IntMan,Ban}),
and (b) the constraint (\ref{restrDxpz}) (or its multidimensional
generalizations \cite{DOM85,167}) can be violated. This does not mean that
the state $\hat{\rho}_{\mathcal{S}}(t)$ can become unphysical -- no, simply
in this case the system $\mathcal{S}$ does not pass over \emph{all possible\/%
} mixed states in the process of evolution, but it moves only along some
specific trajectories in the Hilbert space; see in this connection also \cite%
{Strat97,haake,Diosi93}.

Time-independent diffusion and drift matrices appear only \emph{%
asymptotically\/}, as $t\rightarrow \infty $ (physically, after some
characteristic time determined by the properties of $\mathcal{R}$).
Moreover, the sets of diffusion coefficients satisfying (\ref{restrDxpz})
can be obtained \emph{only for specific forms of the }$\mathcal{R-S}$ \emph{%
interaction Hamiltonian\/}. For example, in the case of a thermal
reservoir, the relaxation superoperator (\ref{ssme1}) can be derived
if the interaction only has the so called \emph{rotating wave
approximation\/} form, which results in the drift matrix (\ref{A})
with $\mu =0$ \cite{DJRLR95}. The superoperator (\ref{ssme2}), with
\emph{arbitrary\/} time independent diffusion coefficients, can be
derived with the use of \emph{squeezed\/} (or \emph{rigged\/})
reservoirs \cite{GaCol85,Duper87}. Our main goal in this paper is to
find the sets of drift and diffusion coefficients (possibly,
time-dependent) that can be obtained from the scheme of quantization
of classical equations with \emph{two\/} stochastic forces.

\section{From classical stochastic forces to quantum master equations}

\label{section-EFME}


\subsection{Classical treatment}

A typical equation of motion of a particle of mass $m$ subjected to a linear
friction force in one dimension is
\begin{equation}
\ddot{x}+\dot{\Gamma}_{t}\dot{x}+\frac{1}{m}\frac{\partial V(x,t)} {\partial
x}= 0,  \label{newton2}
\end{equation}%
where $V(x,t)$ is a potential and $\Gamma _{t}$ a TD dissipative function.
Although Eq. (\ref{newton2}) describes a nonconservative system, it is known
for a long time (see, e.g., \cite{havas,ambig} and references therein) that
it can be derived from some Lagrangian. The most simple one is known under
the name Bateman--Caldirola--Kanai Lagrangian \cite{Bate31,Cal},
\begin{equation}
L\left( x,\dot{x},t\right) =\left( \frac{1}{2}{m\dot{x}^{2}}-V(x,t) \right)
e^{\Gamma _{t}}.  \label{BCK}
\end{equation}%
We consider a simple generalization of (\ref{BCK}),
\begin{equation}
L\left( x,\dot{x},t\right) =\left( \frac{1}{2}{m\dot{x}^{2}}-V(x,t)+xF_{t}+%
\dot{x}G_{t}\right) e^{\Gamma _{t}},  \label{Lagrangian}
\end{equation}%
where $F_{t}$ and $G_{t}$ are arbitrary generalized TD forces associated to
the position $x$ and to the velocity $\dot{x}$.  The \textit{canonical }
momentum is
\begin{equation}
P\equiv {\partial L}/{\partial \dot{x}}=e^{\Gamma _{t}}\left( m\dot{x}%
+G_{t}\right)  \label{mom-can}
\end{equation}%
and we define the \textit{physical} momentum as
\begin{equation}
p\equiv m\dot{x}+G_{t}=Pe^{-\Gamma _{t}}.  \label{mom-mec}
\end{equation}

The Hamiltonian associated to the Lagrangian (\ref{Lagrangian}) is
\begin{eqnarray}
H\left( t\right) &=& P\dot{x}-L =\frac{P^{2}}{2m}e^{-\Gamma _{t}}+\left[
V(x,t)-xF_{t}\right] e^{\Gamma _{t}}  \notag \\
&& -{PG_{t}}/{m} + G_{t}^{2}\exp (\Gamma _{t})/2m.  \label{HP2f}
\end{eqnarray}
The classical equations of motion for canonical coordinates (Hamilton
equations) are
\begin{eqnarray*}
\dot{x} &=&{\partial }H\left( t\right) /{\partial P}=\left( {P}e^{-\Gamma
_{t}}-G_{t}\right) /{m}, \\
\dot{P} &=&-{\partial }H\left( t\right) /{\partial x}=\left( -{\partial
V(x,t)}/{\partial x}+F_{t}\right) e^{\Gamma _{t}},
\end{eqnarray*}%
whereas for the physical coordinates one obtains
\begin{equation*}
\dot{x}={p}/{m} - G_{t}/m, \qquad \dot{p}=-{\partial V(x,t)}/{\partial x}%
+F_{t}-p\dot{\Gamma}_{t},
\end{equation*}%
or as a single second order equation (Newton equation)
\begin{equation}
\ddot{x}+\dot{\Gamma}_{t}\dot{x}+\frac{1}{m}\frac{\partial V(x,t)}{\partial x%
}=\frac{1}{m}\left( F_{t}-\dot{\Gamma}_{t}G_{t}-\dot{G}_{t}\right) ,
\label{newton1}
\end{equation}%
where the RHS contains TD terms only.

We see that in classical mechanics, where the coordinate $x$ is the only
independent variable (since $p$ or $P$ are functions of $\dot{x}$), the
presence of two terms, $-xF_{t}e^{\Gamma _{t}}$ and $-PG_{t}/m$, in
Hamiltonian (\ref{HP2f}) is redundant, because the dynamics depends only on
the combination $\mathcal{F}(t) = F_{t}-\dot{\Gamma}_{t}G_{t}-\dot{G}_{t}$.
A usual choice is $G_t \equiv 0$ and $F_t =\mathcal{F}(t)$. But one can
obtain the same dynamics, choosing $F_t \equiv 0$ and finding the function $%
G_t$ from the equation $-(\dot{\Gamma}_{t}G_{t}+\dot{G}_{t})= {\mathcal{F}}%
(t)$, whose solution is
\begin{equation*}
G_{t}\equiv Ke^{-\Gamma _{t}}-e^{-\Gamma _{t}}\int^{t}e^{\Gamma _{\tau}} {%
\mathcal{F}}(\tau)d\tau,
\end{equation*}%
where $K$ is an arbitrary constant.

However, both the forces, $F_{t}$ and $G_{t}$, are important in  the quantum
case, because of the non-commutation property of position and momentum.
These forces give different contributions to the dynamical evolution of the
system state, as we show in the following subsection.

\subsection{Quantum treatment}

Having the Hamiltonian function (\ref{HP2f}), one can try to
\textquotedblleft quantize\textquotedblright\ the classical dissipative
system, transforming (\ref{HP2f}) to an operator by means of the usual rules
and writing the time-dependent Schr\"{o}dinger equation with this
\textquotedblleft quantum\textquotedblright\ Hamilton operator. This idea
was formulated for the first time by Caldirola and Kanai (CK) in 1940s \cite%
{Cal}, and since that time it was developed or criticized by many authors
(see, e.g., Refs. \cite{Dek,kerner}. It is known by now that such a
simplified approach suffers from many drawbacks. For instance, the CK
Hamiltonian is explicitly time-dependent, so it is closer to the system with
time-dependent mass than to the genuine dissipative system. Moreover, the
problem of finding the Hamiltonian for the given equations of motion has no
unique solution, and practically all such Hamiltonians have some pathology
\cite{havas,ambig}. But the main physical defect of the CK scheme is that it
implies that the quantum state of the system remains \emph{pure\/} during
the evolution, because regular classical fields interacting with a quantum
system $\mathcal{S}$ do not change its informational content as time goes
on, even if energy is not conserved; an initial pure state $\psi _{\mathcal{S%
}}(0)$ will evolves as a pure state $\psi _{\mathcal{S}}(t)$. On the other
hand, it is known that dissipation is connected with a loss of quantum
purity. Thus one has to describe the system in terms of the density matrix
or its equivalent forms, such as the Wigner function, for example. But how
to find equations of motion for the density matrix?

An answer was given in Refs. \cite{Pap73,svinin,dekker2,brimiz,AlMes82}: one
should start not from the Schr\"{o}dinger equation for a wave function, but
from the von Neumann--Liouville equation for the statistical operator,
considering $F_{t}$ and $G_{t}$ as \emph{stochastic forces\/} and performing
some averaging over these forces. This averaging results in an information
drain from the quantum system $\mathcal{S}$. So irreversibility is verified
and the entropy of the system changes due to the random character of the
classical fields. However, by using a \emph{single\/} stochastic force one
obtains equations which not always preserve the property of positivity of
the statistical operator. Our goal here is to show that correct equations,
satisfying all principles of QM, can be derived in a very simple way, if one
introduces \emph{two\/} classical stochastic forces, instead of a single
one, besides one regular dissipative function. Such a derivation was not
done earlier, as far as we know.

The equation for the time evolution of the density matrix following from
Hamiltonian (\ref{HP2f}) reads
\begin{equation}
\frac{d\hat{\rho}_{t}}{dt}=\frac{1}{i\hbar }\left[ \hat{H}_{ef}(t) +\hat{W}(
\hat{x},\hat{P},t), \hat{\rho}_{t}\right] ,  \label{liouville2}
\end{equation}%
where
\begin{equation}
\hat{H}_{ef}\left( t\right) =e^{-\Gamma _{t}}{\hat{P}^{2}}/(2m)+V\left( \hat{%
x},t\right) e^{\Gamma _{t}},  \label{kanai}
\end{equation}
\begin{equation}
\hat{W}( \hat{x},\hat{P},t) = -e^{\Gamma _{t}}\hat{x}F_{t}-{\hat{P}}G_{t}/m.
\label{W}
\end{equation}%
$\hat{W}( \hat{x},\hat{P},t) $ is a stochastic operator if $F_{t}$ and $G_{t}
$ are assumed as stochastic forces. The TD term $G_{t}^{2}\exp (\Gamma
_{t})/2m$ was thrown out from $\hat{H}_{ef}\left( t\right)$ since it does
not contribute to the equations of motion.

Using the unitary evolution operator $\hat{U}_{t}$ corresponding to the
effective free Hamiltonian (\ref{kanai}),
\begin{equation*}
i\hbar \,{d\hat{U}_{t}}/{dt}=\hat{H}_{ef}\left( t\right) \hat{U}_{t},
\end{equation*}%
we make a unitary transformation
\begin{equation}
\hat{\rho}_{t}=\hat{U}_{t}\tilde{\rho}_{t}\hat{U}_{t}^{\dagger },  \label{A4}
\end{equation}%
which removes the term $\hat{H}_{ef}\left( t\right) $ from Eq. (\ref%
{liouville2}):
\begin{equation}
i\hbar \,{d\tilde{\rho}}_{t}/{dt}=\left[ {\hat{W}}(\tilde{x}_{t},\tilde{P}%
_{t},t),\tilde{\rho}_{t}\right] \equiv \left[ \tilde{W}(t),\tilde{\rho}_{t}%
\right] ,  \label{A12}
\end{equation}%
and where%
\begin{equation}
\tilde{x}_{t}=U_{t}^{\dagger }\hat{x}U_{t},\qquad \tilde{P}%
_{t}=U_{t}^{\dagger }\hat{P}U_{t}.  \label{A13a}
\end{equation}%
A formal solution to Eq. (\ref{A12}) is
\begin{equation}
\tilde{\rho}_{t}=\hat{\rho}_{0}+\frac{1}{i\hbar }\int_{0}^{t}dt^{\prime }%
\left[ \tilde{W}(t^{\prime }),\tilde{\rho}_{t^{\prime }}\right] .
\label{A13}
\end{equation}%
Iterating Eq. (\ref{A13}) and deriving with respect to time we get the
equation%
\begin{eqnarray}
\frac{d\tilde{\rho}_{t}}{dt} &=&\frac{1}{i\hbar }\left[ {\tilde{W}}(t),\hat{%
\rho}_{0}\right]   \notag \\
&&+\frac{1}{\left( i\hbar \right) ^{2}}\int_{0}^{t}dt^{\prime }\left[ \tilde{%
W}\left( t\right) ,\left[ \tilde{W}\left( t^{\prime }\right) ,\tilde{\rho}%
_{t^{\prime }}\right] \right]   \label{A13d}
\end{eqnarray}%
which is still exact. Its physical content is the same as Eq. (\ref{A12}).
Using the RHS of (\ref{A13}) for $\tilde{\rho}_{t^{\prime }}$ and inserting
it in (\ref{A13d}) recursively, we obtain an infinite series
\begin{eqnarray}
\frac{d\tilde{\rho}_{t}}{dt} &=&\frac{1}{i\hbar }\left[ \tilde{W}(t),\hat{%
\rho}_{0}\right] +\sum_{k=1}^{\infty }\frac{1}{(i\hbar )^{k+1}}%
\int_{0}^{t}dt_{1}\cdots \int_{0}^{t_{k-1}}dt_{k}  \notag \\
&&\times \left[ \tilde{W}(t),\left[ \tilde{W}(t_{1}),\cdots \left[ \tilde{W}%
(t_{k}),\hat{\rho}_{0}\right] \right] \right] .  \label{ser-k}
\end{eqnarray}%
The next step is to perform averaging over stochastic forces in Eq. (\ref%
{ser-k}). We assume that classical stochastic functions $F_{t}$ and $G_{t}$
are Gaussian with zero average values, $\overline{F_{t}}=\overline{G_{t}}=0$%
, and that they are delta-correlated with a TD function (not a stationary
process):
\begin{eqnarray}
\overline{F_{t_{1}}F_{t_{2}}} &=&2A_{t_{1}}\delta \left( t_{1}-t_{2}\right) ,
\label{A12i1} \\
\overline{G_{t_{1}}G_{t_{2}}} &=&2m^{2}B_{t_{1}}\delta \left(
t_{1}-t_{2}\right) ,  \label{A12i2} \\
\overline{F_{t_{1}}G_{t_{2}}} &=&2mC_{t_{1}}\delta \left( t_{1}-t_{2}\right)
.  \label{A12i3}
\end{eqnarray}%
Moreover, we take into account the important properties of \textit{Gaussian}%
\emph{\/} stochastic processes, namely, $\overline{%
J_{t_{1}}J_{t_{2}}...J_{t_{2n+1}}}=0$ for an odd number of terms ($J_{t}$
stands for $F_{t}$ or $G_{t}$), whereas for an even number
\begin{equation}
\overline{J_{t_{1}}J_{t_{2}}...J_{t_{2n}}}=\sum_{all\text{ }pairs}\overline{%
J_{t_{i}}J_{t_{j}}}\,\cdot \,\overline{J_{t_{k}}J_{t_{l}}},  \label{A12c}
\end{equation}%
where the average is over ensembles. Since stochastic operators ${\tilde{W}}%
(t_{k})$ are linear combinations of $\tilde{x}_{t_{k}}$ and $\tilde{P}%
_{t_{k}}$, only the terms with even numbers of operators $\tilde{W}(t_{k})$
survive after the averaging in Eq. (\ref{ser-k}), so that we arrive at the
series containing only even powers of $\hbar $:
\begin{equation}
\frac{{d\hat{\rho}}_{t}}{{dt}}=\sum_{k=1}^{\infty }\frac{1}{(i\hbar )^{2k}}%
\hat{\chi}_{2k}\left( t\right) .  \label{ser-2}
\end{equation}%
The first term of this expansion is a sum of four integrals containing
double commutators,
\begin{eqnarray*}
\hat{\chi}_{2}\left( t\right)  &=&\int_{0}^{t}dt^{\prime }e^{\Gamma
_{t}+\Gamma _{t^{\prime }}}\overline{F_{t}F_{t^{\prime }}}\left[ \tilde{x}%
_{t},\left[ \tilde{x}_{t^{\prime }},\hat{\rho}_{0}\right] \right]  \\
&&+{m^{-1}}e^{\Gamma _{t}}\int_{0}^{t}dt^{\prime }\overline{%
F_{t}G_{t^{\prime }}}\left[ \tilde{x}_{t},\left[ \tilde{P}_{t^{\prime }},%
\hat{\rho}_{0}\right] \right]  \\
&&+{m^{-1}}\int_{0}^{t}dt^{\prime }e^{\Gamma _{t^{\prime }}}\overline{%
G_{t}F_{t^{\prime }}}\left[ \tilde{P}_{t},\left[ \tilde{x}_{t^{\prime }},%
\hat{\rho}_{0}\right] \right]  \\
&&+{m^{-2}}\int_{0}^{t}dt^{\prime }\overline{G_{t}G_{t^{\prime }}}\left[
\tilde{P}_{t},\left[ \tilde{P}_{t^{\prime }},\hat{\rho}_{0}\right] \right] ,
\end{eqnarray*}%
which can be easily calculated due to the presence of delta-functions in
Eqs. (\ref{A12i1})--(\ref{A12i3}), resulting in the following expression:
\begin{eqnarray}
\hat{\chi}_{2}\left( t\right)  &=&A_{t}e^{2\Gamma _{t}}\left[ \tilde{x}_{t},%
\left[ \tilde{x}_{t},\hat{\rho}_{0}\right] \right] +B_{t}\left[ \tilde{P}%
_{t},\left[ \tilde{P}_{t},\hat{\rho}_{0}\right] \right]   \notag \\
&&+C_{t}e^{\Gamma _{t}}\left( \left[ \tilde{x}_{t},\left[ \tilde{P}_{t},\hat{%
\rho}_{0}\right] \right] +\left[ \tilde{P}_{t},\left[ \tilde{x}_{t},\hat{\rho%
}_{0}\right] \right] \right) .  \label{A14}
\end{eqnarray}%
Continuing these steps, we see that the structure of the term $\hat{\chi}%
_{2}\left( t\right) $ is repeated each time, resulting finally in replacing
the initial operator $\hat{\rho}_{0}$ by the time dependent operator $\hat{%
\rho}_{t}$. Thus we obtain the following closed equation governing the time
evolution of the statistical operator averaged over stochastic forces (see
Appendix A for details of the derivation):
\begin{eqnarray}
\frac{d\hat{\rho}_{t}}{dt} &=&\frac{1}{i\hbar }\left[ \hat{H}_{ef}(t),\hat{%
\rho}_{t}\right] -\frac{1}{\hbar ^{2}}\left( A_{t}e^{2\Gamma _{t}}\left[
\hat{x},\left[ \hat{x},\hat{\rho}_{t}\right] \right] \right.   \notag \\
&&\left. +B_{t}[\hat{P},[\hat{P},\hat{\rho}_{t}]]+2C_{t}e^{\Gamma _{t}}[\hat{%
x},[\hat{P},\hat{\rho}_{t}]]\right) .  \label{2pme}
\end{eqnarray}%
We would like to emphasize that no truncations of higher order terms were
done in deriving Eq. (\ref{2pme}), so this equation holds for any TD force
strengths (coefficients $A_{t},B_{t},C_{t}$). A possible additional term
proportional to $[\hat{P},[\hat{x},\hat{\rho}_{t}]]$ in the RHS of Eq. (\ref%
{2pme}) is redundant because of the identity $[\hat{x},[\hat{P},\hat{\rho}%
_{t}]]=[\hat{P},[\hat{x},\hat{\rho}_{t}]]$. Had we assumed one force $F_{t}$
only, the coefficients $B_{t}$ and $C_{t}$ would be zero. Equation (\ref%
{2pme}) is structurally analogous to that obtained by Hu, Paz and Zhang \cite%
{Hu92}, although their TD coefficients were derived assuming a
reservoir made of HO's, while ours are purely phenomenological.
Below we show that the associated Wigner functions coincide, and as
such Eq. (\ref{2pme}) contains the non-Markovian effects (in the
sense of Refs. \cite{haake,Hu92,IntMan,Ban}, i.e., time-dependent
diffusion and drift coefficients). In our derivation, limits as high
or low-temperature, strong or weak-coupling have no room for
discussion since all the effects of the environment on the system
depend on the adopted values for the four TD parameters. In
particular, for coefficients $A$, $B$ and $C$ being time-independent
one retrieves the Markovian limit.

Eq. (\ref{2pme}) cannot be immediately identified with Eq. (\ref{ssme2}),
due to different meanings of the variables $\hat{p}$ (physical momentum) in (%
\ref{ssme2}) and $\hat{P}$ (canonical momentum) in (\ref{2pme}), besides the
presence of the factors\ $\exp (\Gamma _{t})$ and $\exp (2\Gamma _{t})$.
However, it is easy to show that these equations are physically equivalent,
because they give the same Wigner function for the mapped physical
coordinates.

The equations of motion of first and second moments for the canonical
variables are (the average values are defined as ${\langle \hat{A}\rangle }{%
\equiv }$ Tr$(\hat{A}\hat{\rho}_{t})$)
\begin{equation*}
{d\left\langle \hat{x}\right\rangle }/{dt}=e^{-\Gamma _{t}}{\langle \hat{P}%
\rangle }/{m},\qquad {d\langle \hat{P}\rangle }/{dt}=\left\langle -{\partial
V}/{\partial {x}}\right\rangle e^{\Gamma _{t}},
\end{equation*}%
\begin{equation*}
{d\left\langle \hat{x}^{2}\right\rangle }/{dt}=e^{-\Gamma _{t}}\langle \{%
\hat{x},\hat{P}\}\rangle /m+2B_{t},
\end{equation*}%
\begin{equation*}
{d\langle \hat{P}^{2}\rangle }/{dt}=-\langle \{\hat{P},{\partial \hat{V}}/{%
\partial x}\}\rangle e^{\Gamma _{t}}+2A_{t}e^{2\Gamma _{t}},
\end{equation*}%
\begin{equation*}
\frac{d}{dt}\langle \{\hat{x},\hat{P}\}\rangle =\frac{2}{m}\langle \hat{P}%
^{2}\rangle e^{-\Gamma _{t}}-2\langle \hat{x}\frac{\partial \hat{V}}{%
\partial x}\rangle e^{\Gamma _{t}}+4C_{t}e^{\Gamma _{t}},
\end{equation*}%
and one notices that these equations contain the TD exponential factors $%
\exp (\pm \Gamma _{t})$. However, passing to the physical momentum (\ref%
{mom-mec}), we get rid of these factors,
\begin{equation}
{d\left\langle \hat{x}\right\rangle }/{dt}={\left\langle \hat{p}%
\right\rangle }/{m},  \label{mov1x}
\end{equation}%
\begin{equation}
{d\left\langle \hat{p}\right\rangle }/{dt}=-\langle {\partial \hat{V}}/{%
\partial \hat{x}}\rangle -\dot{\Gamma}_{t}\left\langle \hat{p}\right\rangle ,
\label{mov1}
\end{equation}%
\begin{equation}
{d\left\langle \hat{x}^{2}\right\rangle }/{dt}=m^{-1}\left\langle \left\{
\hat{x},\hat{p}\right\} \right\rangle +2B_{t},  \label{mov2}
\end{equation}%
\begin{equation}
{d\left\langle \hat{p}^{2}\right\rangle }/{dt}=-\langle \{\hat{p},{\partial
\hat{V}}/{\partial \hat{x}}\}\rangle -2\dot{\Gamma}_{t}\left\langle \hat{p}%
^{2}\right\rangle +2A_{t},  \label{mov3}
\end{equation}%
\begin{equation}
\frac{d}{dt}\left\langle \left\{ \hat{x},\hat{p}\right\} \right\rangle =%
\frac{2}{m}{\left\langle \hat{p}^{2}\right\rangle }-2\langle x\frac{\partial
\hat{V}}{\partial \hat{x}}\rangle -\dot{\Gamma}_{t}\langle \left\{ \hat{x},%
\hat{p}\right\} \rangle +4C_{t}.  \label{mov4}
\end{equation}%
Comparing Eqs. (\ref{A}) and (\ref{dot-q-A}) with (\ref{mov1x}) and (\ref%
{mov1}), we see that they coincide if $\hat{V}(x)=m\omega _{0}^{2}\hat{x}%
^{2}/2$, $\Gamma _{t}=2\lambda t$, and $\mu =\lambda $. Then comparing
equations for the covariances of the physical momentum and coordinate, one
can verify that they satisfy the matrix equation (\ref{wigner3}) if the
diffusion coefficients are identified as
\begin{equation}
A_{t}=D_{x}\left( t\right) ,\quad B_{t}=D_{p}\left( t\right) ,\quad
C_{t}=D_{z}\left( t\right) .  \label{con-PME}
\end{equation}

In what follows we will show the equivalence between the Wigner function $%
W^{P}(\mathbf{Q},t)$ in the \emph{canonical\/} phase space, $\mathbf{Q}=(x,P)
$ with the Wigner function in the physical variables phase space $W(\mathbf{q%
},t)$. The $W^{P}(\mathbf{Q},t)$ is governed by the Fokker--Planck equation (%
\ref{wigner}) with time-dependent drift and diffusion matrices,
\begin{equation*}
\mathbf{A}^{P}=\left\Vert
\begin{array}{cc}
0 & {e^{-\lambda t}}/{m} \\
-m\omega _{0}^{2}e^{\lambda t} & 0%
\end{array}%
\right\Vert ,\quad \mathbf{D}^{P}=\left\Vert
\begin{array}{cc}
B_{t} & C_{t}e^{\lambda t} \\
C_{t}e^{\lambda t} & A_{t}e^{2\lambda t}%
\end{array}%
\right\Vert .
\end{equation*}%
The solution of Eq. (\ref{wigner}) for the function $W(\mathbf{Q},t)$ can be
written as
\begin{equation*}
W^{P}(\mathbf{{Q},}t\mathbf{)=\int }G^{P}\mathbf{({Q},{Q}^{\prime },}t%
\mathbf{)}W^{P}\mathbf{({Q}^{\prime },}0\mathbf{)}d\mathbf{{Q}^{\prime }},
\end{equation*}%
where the propagator is given by the formula \cite{Agar73,Dbook03}
\begin{eqnarray}
&&G^{P}(\mathbf{{Q},{Q^{\prime }},}t)=\left( 2\pi \sqrt{\det \mathbf{N}(t)}%
\right) ^{-1}  \notag \\
&&\times \exp \left[ -\frac{1}{2}(\mathbf{Q}-\mathbf{R}^{P}(t)\mathbf{Q}%
^{\prime })\mathbf{N}^{-1}(t)(\mathbf{Q}-\mathbf{R}^{P}(t)\mathbf{Q}^{\prime
})\right] .  \notag \\
&&  \label{green4}
\end{eqnarray}%
Matrix $\mathbf{N}(t)$ satisfies Eq. (\ref{wigner3}) (with matrices $\mathbf{%
A}^{P}$ and $\mathbf{D}^{P}$) and the initial condition $\mathbf{N}(0)=0$,
whereas matrix $\mathbf{R}^{P}(t)$ satisfies the equation ${d\mathbf{R}^{P}}/%
{dt}=\mathbf{A}^{P}(t)\mathbf{R}^{P}$ and the initial condition $\mathbf{R}%
^{P}(0)=\mathbf{1}$ (unity matrix).

Differential equations for three different matrix elements of the
symmetrical matrix $\mathbf{N}\left( t\right) $ have the form
\begin{eqnarray*}
dN_{11}/dt &=&2e^{-\lambda t}N_{12}/m+2B_{t}, \\
dN_{12}/dt &=&e^{-\lambda t}N_{22}/m-m\omega _{0}^{2}e^{\lambda
t}N_{11}+2C_{t}e^{\lambda t}, \\
dN_{22}/dt &=&-2m\omega _{0}^{2}e^{\lambda t}N_{12}+2A_{t}e^{2\lambda t}.
\end{eqnarray*}%
Doing the substitution (\ref{con-PME}) and the changes
\begin{equation*}
N_{22}=e^{2\lambda t}M_{22},\quad N_{12}=e^{\lambda t}M_{12},\quad
N_{11}=M_{11}
\end{equation*}%
we obtain the equations for elements of matrix $\mathbf{M}\left(
t\right) $ with \textit{time-independent}\emph{\/} drift and
diffusion matrices, given by Eqs. (\ref{A}) and (\ref{wigner3}) with
$\lambda =\mu $. Then one can verify that the propagators in the
canonical and physical phase spaces are related by a simple formula
\begin{equation}
G^{P}(\mathbf{{Q},{Q^{\prime }},}t\mathbf{)}=e^{-\lambda t}G(\mathbf{{q},{%
q^{\prime }},}t\mathbf{).}  \label{green}
\end{equation}%
Therefore, as the initial Wigner function is the same in both coordinate
systems, $W^{P}\left( \mathbf{Q},0\right) =W\left( \mathbf{q},0\right) $,
the Wigner function at any time becomes
\begin{eqnarray}
W^{P}(\mathbf{Q},t) &=&\int G^{P}(\mathbf{Q};\mathbf{Q}^{\prime },t)W^{P}(%
\mathbf{Q}^{\prime },0)\,d\,\mathbf{Q}^{\prime }  \notag \\
&=&\int e^{-\lambda t}G(x,p;x^{\prime },p^{\prime },t)W(x^{\prime
},p^{\prime };0)e^{\lambda t}dx^{\prime }dp^{\prime }  \notag \\
&=&\int G(x,p;x^{\prime },p^{\prime },t)W(x^{\prime },p^{\prime
};0)\,dx^{\prime }dp^{\prime }\   \notag \\
&=&W(\mathbf{q},t).
\end{eqnarray}%
Thus the master equation (\ref{2pme}) is completely equivalent to the
translationally invariant ($\lambda =\mu $) form of the master equation for
quantum Brownian motion (\ref{ssme1}) with operators (\ref{Hmu}) and (\ref%
{ssme2}).


\section{Summary and conclusions}

\label{section-summary}


We have analyzed the phenomenological approach to build a master equation
for describing the irreversible and dissipative dynamical evolution of the
state of a quantum system $\mathcal{S}$, under the influence of an
unspecified environment. In contradistinction to the microscopic approach
that models the environment as a reservoir $\mathcal{R}$ made of an infinite
number of degrees of freedom (for example, harmonic oscillators), the
phenomenological approach makes use of dissipative functions and stochastic
forces. We showed that the Newton equation of motion for $\mathcal{S}$ does
not change by introducing \textit{two} such forces instead of only one,
however when we do the quantization of the system, both forces become quite
important, contributing on equal footing. We derived from first principles,
with a Lagrangian containing \textit{one} dissipative function and \textit{%
two} stochastic forces, the master equation describing the quantum Brownian
motion (of a harmonic oscillator) with translationally invariant damping and
the most general bilinear (with respect to the coordinate and momentum
operators) relaxation superoperator, which can be reduced to an equivalent
differential equation of the Fokker--Planck type. However, the TD
phenomenological parameters entering the forces cannot be determined within
the framework of the phenomenological approach; they should be fixed either
from experimental data that reproduce relevant physical properties of $%
\mathcal{S}$ or from some other considerations, such as the requirement of
satisfying the positivity constraint (\ref{restrDxpz}) or by fitting
properties at thermal equilibrium. In this direction we verified that the
master equations derived in \cite{haake,UnZu89,Hu92}, containing
non-Markovian effects (present in their TD coefficients) are accounted in
our derivation where the environment is simulated by two effective forces
and the dissipative function, instead of assuming an interaction with  an
infinite set of HO's.

We would like to emphasize that by averaging over the stochastic forces, we
did not disregard any term (see Appendix for details). In this sense, the
phenomenological derivation of the master equation is as exact as other
approaches \cite{Hu92}. Of course, this happened due to the choice of
correlation functions in the form of delta functions, albeit multiplied by
time-dependent strength factors. The pertinent question, what could happen
in the most general non-Markovian case, when correlation functions, such as $%
A(t_{1},t_{2})$, are arbitrary functions of the time difference $t_{1}-t_{2}$
(colored noise), requires a separate study. Certainly, the phenomenological
approach used in this paper has limitations, because it is based on some
effective Lagrangian. Therefore, although it works well for one-dimensional
systems (or isotropical multidimensional ones), it will fail for generic
multidimensional systems with several independent damping coefficients,
because no effective Lagrangian can be found for such systems \cite%
{havas,ambig}. This explains also why only a subfamily of master equations (%
\ref{ssme2}), restricted by the condition $\lambda =\mu $ (translationally
invariant damping), can be obtained within the framework of the scheme used
in this paper: there are no effective Lagrangians for $\lambda \neq \mu $
[i.e., for \emph{two\/} \textquotedblleft friction forces\textquotedblright\
in the classical equations of motion (\ref{eqmeanx}) and (\ref{eqmeanp})].

\begin{acknowledgments}
AVD and SSM acknowledge financial support from FAPESP, S\~{a}o Paulo
(contract \# 04/13705-3). SSM and VVD acknowledge financial support from
CNPq, Brasil.
\end{acknowledgments}

%
\appendix
%

\section{Derivation of the Phenomenological Master Equation}

\label{apendiceA}

Iterating Eq. (\ref{A13}) once more we get the formal solution
\begin{eqnarray}
\tilde{\rho}_{t} &=&\hat{\rho}_{0}+\frac{1}{i\hbar }\int_{0}^{t}dt_{1}\left[
\tilde{W}(t_{1}),\hat{\rho}_{0}\right]   \label{A13c} \\
&+&\left( \frac{1}{i\hbar }\right)
^{2}\int_{0}^{t}dt_{1}\int_{0}^{t_{1}}dt_{2}\left[ \tilde{W}\left(
t_{1}\right) ,\left[ \tilde{W}\left( t_{2}\right) ,\hat{\rho}_{t_{2}}\right] %
\right] ,  \notag
\end{eqnarray}%
and deriving it with respect to time we get Eq. (\ref{A13d}). Using the RHS
of (\ref{A13c}) for $\tilde{\rho}_{t^{\prime }}$, inserting it in the RHS of
Eq. (\ref{A13d}) and keeping even terms in $\tilde{W}$ (because only the
even terms will survive after averaging over the ensemble), Eq. (\ref{A13d})
may be written as an infinite series
\begin{eqnarray*}
\frac{d\tilde{\rho}_{t}}{dt} &=&\frac{1}{\left( i\hbar \right) ^{2}}%
\int_{0}^{t}dt^{\prime }\Big[\tilde{W}\left( t\right) ,\Big[\tilde{W}\left(
t^{\prime }\right) ,\hat{\rho}_{0}+\frac{1}{\left( i\hbar \right) ^{2}} \\
&&\times \int_{0}^{t^{\prime }}dt_{1}\int_{0}^{t_{1}}dt_{1}^{\prime }\left[
\tilde{W}\left( t_{1}\right) ,\left[ \tilde{W}\left( t_{1}^{\prime }\right) ,%
\hat{\rho}_{0}\right] \right] +\cdots \Big]\Big]
\end{eqnarray*}%
or
\begin{eqnarray}
\frac{d\tilde{\rho}_{t}}{dt} &=&\frac{1}{\left( i\hbar \right) ^{2}}%
\int_{0}^{t}dt^{\prime }\left[ \tilde{W}\left( t\right) ,\left[ \tilde{W}%
\left( t^{\prime }\right) ,\hat{\rho}_{0}\right] \right]   \notag \\
&&+\frac{1}{\left( i\hbar \right) ^{2}}\int_{0}^{t}dt^{\prime }\left[ \tilde{%
W}\left( t\right) ,\left[ \tilde{W}\left( t^{\prime }\right)
,\sum_{k=1}^{\infty }\frac{1}{\left( i\hbar \right) ^{2k}}\right. \right.
\notag \\
&&\left. \left. \times \int_{0}^{t^{\prime
}}dt_{1}\int_{0}^{t_{1}}dt_{1}^{\prime }\ldots \int_{0}^{t_{k-1}^{\prime
}}dt_{k}\int_{0}^{t_{k}}dt_{k}^{\prime }\right. \right.   \notag \\
&&\left. \left. \times \left[ \tilde{W}\left( t_{1}\right) ,\left[ \tilde{W}%
\left( t_{1}^{\prime }\right) ,...\left[ \tilde{W}\left( t_{k}\right) ,\left[
\tilde{W}\left( t_{k}^{\prime }\right) ,\hat{\rho}_{0}\right] \right] \right]
\right] \right] \right] ,  \notag \\
&&  \label{A13g}
\end{eqnarray}%
with $t_{0}^{\prime }\equiv t^{\prime }$. Averaging the quadratic term in $%
\tilde{W}\left( \cdot \right) $ of (\ref{A13g}), we get the four terms of
Eq. (\ref{A14}). Averaging the quartic terms in $\tilde{W}\left( \cdot
\right) $,
\begin{eqnarray*}
&&\frac{1}{\left( i\hbar \right) ^{4}}\int_{0}^{t}dt^{\prime
}\int_{0}^{t^{\prime }}dt_{1}\int_{0}^{t_{1}}dt_{1}^{\prime } \\
&&\times \overline{\left[ \tilde{W}\left( t\right) ,\left[ \tilde{W}\left(
t^{\prime }\right) ,\left[ \tilde{W}\left( t_{1}\right) ,\left[ \tilde{W}%
\left( t_{1}^{\prime }\right) ,\hat{\rho}_{0}\right] \right] \right] \right]
},
\end{eqnarray*}%
we obtain sixteen terms, which together with the four terms of (\ref{A14})
give the expression
\begin{eqnarray*}
&&\frac{1}{\left( i\hbar \right) ^{2}}\left\{ A_{t}e^{2\Gamma _{t}}\left[
\tilde{x}\left( t\right) ,\left[ \tilde{x}\left( t\right) ,\tilde{\rho}%
_{t^{\prime }}\right] \right] \right.  \\
&&\left. +C_{t}te^{\Gamma _{t}}\left( \left[ \tilde{x}\left( t\right) ,\left[
\tilde{P}\left( t\right) ,\tilde{\rho}_{t^{\prime }}\right] \right] +\left[
\tilde{P}\left( t\right) ,\left[ \tilde{x}\left( t\right) ,\tilde{\rho}%
_{t^{\prime }}\right] \right] \right) \right.  \\
&&\left. +B_{t}\left[ \tilde{P}\left( t\right) ,\left[ \tilde{P}\left(
t\right) ,\tilde{\rho}_{t^{\prime }}\right] \right] \right\} ,
\end{eqnarray*}%
where%
\begin{eqnarray}
\tilde{\rho}_{t^{\prime }} &=&\hat{\rho}_{0}+\frac{1}{\left( i\hbar \right)
^{2}}\int_{0}^{t^{\prime }}dt^{\prime \prime }\ \left\{ A_{t^{\prime \prime
}}e^{2\Gamma _{t^{\prime \prime }}}\left[ \tilde{x}_{t^{\prime \prime }},%
\left[ \tilde{x}_{t^{\prime \prime }},\hat{\rho}_{0}\right] \right] \right.
\notag \\
&&\left. +C_{t^{\prime \prime }}e^{\Gamma _{t^{\prime \prime }}}\left( \left[
\tilde{x}_{t^{\prime \prime }},\left[ \tilde{P}_{t^{\prime \prime }},\hat{%
\rho}_{0}\right] \right] +\left[ \tilde{P}_{t^{\prime \prime }},\left[
\tilde{x}_{t^{\prime \prime }},\hat{\rho}_{0}\right] \right] \right) \right.
\notag \\
&&\left. +B_{t^{\prime \prime }}\left[ \tilde{P}_{t^{\prime \prime }},\left[
\tilde{P}_{t^{\prime \prime }},\hat{\rho}_{0}\right] \right] \right\}
+\cdots .  \label{A16}
\end{eqnarray}%
Finally, after averaging and collecting all terms in Eq. (\ref{A13g}) one
obtains the master equation in the interaction picture%
\begin{eqnarray}
\frac{d\tilde{\rho}_{t}}{dt} &=&\frac{1}{\left( i\hbar \right) ^{2}}\left\{
A_{t}e^{2\Gamma _{t}}\left[ \tilde{x}_{t},\left[ \tilde{x}_{t},\tilde{\rho}%
_{t}\right] \right] +B_{t}\left[ \tilde{P}_{t},\left[ \tilde{P}_{t},\tilde{%
\rho}_{t}\right] \right] \right.   \notag \\
&&\left. +C_{t}e^{\Gamma _{t}}\left( \left[ \tilde{x}_{t},\left[ \tilde{P}%
_{t},\tilde{\rho}_{t}\right] \right] +\left[ \tilde{P}_{t},\left[ \tilde{x}%
_{t},\tilde{\rho}_{t}\right] \right] \right) \right\} ,  \label{Aeqmaster1}
\end{eqnarray}%
which is exact. Note that on the RHS of Eq. (\ref{A16}) the density operator
is time-independent whereas it is TD in Eq. (\ref{Aeqmaster1}).

To illustrate the calculations, let us consider an example of quartic terms
of the form
\begin{eqnarray*}
&&\frac{1}{\left( i\hbar \right) ^{4}}\int_{0}^{t}dt^{\prime }\
\int_{0}^{t^{\prime }}dt^{\prime \prime }\int_{0}^{t^{\prime \prime
}}dt^{\prime \prime \prime }\exp \left( \Gamma _{t}+\Gamma _{t^{\prime
}}+\Gamma _{t^{\prime \prime }}+\Gamma _{t^{\prime \prime \prime }}\right)
\\
&&\times \overline{F_{t}F_{t^{\prime }}F_{t^{\prime \prime }}F_{t^{\prime
\prime \prime }}}\left[ \tilde{x}_{t},\left[ \tilde{x}_{t^{\prime }},\left[
\tilde{x}_{t^{\prime \prime }},\left[ \tilde{x}_{t^{\prime \prime \prime }},%
\hat{\rho}_{0}\right] \right] \right] \right] .
\end{eqnarray*}%
They give the following terms in the master equation:
\begin{eqnarray*}
\frac{d\tilde{\rho}_{t}}{dt} &=&\frac{2}{\left( i\hbar \right) ^{2}}%
\int_{0}^{t}dt_{2}\ e^{\Gamma _{t}+\Gamma _{t_{2}}}A_{t}\delta \left(
t-t_{2}\right) \left[ \tilde{x}_{t},\left[ \tilde{x}_{t_{2}},\hat{\rho}_{0}%
\right] \right]  \\
&&+\frac{2^{2}}{\left( i\hbar \right) ^{4}}\int_{0}^{t}dt_{2}%
\int_{0}^{t_{2}}dt_{3}\int_{0}^{t_{3}}dt_{4}e^{\Gamma _{t}+\Gamma
_{t_{2}}+\Gamma _{t_{3}}+\Gamma _{t_{4}}} \\
&&\times \left[ \tilde{x}_{t},\left[ \tilde{x}_{t_{2}},\left[ \tilde{x}%
_{t_{3}},\left[ \tilde{x}_{t_{4}},\hat{\rho}_{0}\right] \right] \right] %
\right]  \\
&&\times \left\{ A_{t}\delta \left( t-t_{2}\right) A_{t_{3}}\delta \left(
t_{3}-t_{4}\right) \right.  \\
&&\left. +A_{t}\delta \left( t-t_{3}\right) A_{t_{2}}\delta \left(
t_{2}-t_{4}\right) \right.  \\
&&\left. +A_{t_{2}}\delta \left( t_{2}-t_{3}\right) A_{t}\delta \left(
t-t_{4}\right) \right\} +\cdots .
\end{eqnarray*}%
Due to the time ordering $t\geq t_{2}\geq t_{3}\geq t_{4}$, only the
products of delta-functions $\delta \left( t-t_{2}\right) \delta \left(
t_{3}-t_{4}\right) $ contribute to the integral, and in general,
\begin{equation}
\overline{F_{t_{1}}F_{t_{2}}...F_{t_{2n}}}\equiv
2^{n}\prod_{i=1}^{n}A_{t_{2i-1}}\delta \left( t_{2i-1}-t_{2i}\right) .
\label{deltaprod}
\end{equation}%
Thus the terms proportional to the coefficient $A_{t}$ can be combined as
follows,
\begin{eqnarray}
\frac{d\tilde{\rho}_{t}}{dt} &=&\frac{A_{t}}{\left( i\hbar \right) ^{2}}%
e^{2\Gamma _{t}}\left[ \tilde{x}_{t},\left[ \tilde{x}_{t},\hat{\rho}_{0}%
\right] \right] +\frac{2^{2}A_{t}}{\left( i\hbar \right) ^{4}}\left( \frac{1%
}{2}\right) ^{2}e^{2\Gamma _{t}}  \notag \\
&&\times \int_{0}^{t}dt_{3}A_{t_{3}}e^{2\Gamma _{t_{3}}}\left[ \tilde{x}_{t},%
\left[ \tilde{x}_{t},\left[ \tilde{x}_{t_{3}},\left[ \tilde{x}_{t_{3}},\hat{%
\rho}_{0}\right] \right] \right] \right] +\cdots   \notag \\
&=&\frac{A_{t}}{\left( i\hbar \right) ^{2}}e^{2\Gamma _{t}}\left[ \tilde{x}%
_{t},\left[ \tilde{x}_{t},\tilde{\rho}_{t}\right] \right] ,
\label{contrib-A}
\end{eqnarray}%
if one notices that
\begin{equation*}
\tilde{\rho}_{t}=\hat{\rho}_{0}+\frac{1}{\left( i\hbar \right) ^{2}}%
\int_{0}^{t}dt_{3}\ A_{t_{3}}e^{2\Gamma _{t_{3}}}\left[ \tilde{x}_{t_{3}},%
\left[ \tilde{x}_{t_{3}},\hat{\rho}_{0}\right] \right] +\cdots .
\end{equation*}%
The factor $1/2$ in the first line of (\ref{contrib-A}) occurs due to the
integration of the Dirac delta function over a \textit{semi-infinite\/}
interval. We emphasize that Eq. (\ref{contrib-A}) is \textit{exact\/} under
the assumption (\ref{A12i1}), because no terms were disregarded.

Returning to the original operator $\hat{\rho}_{t}$ with the aid of the
transformation (\ref{A4}), we arrive at Eq. (\ref{2pme}).

\end{document}